\documentstyle[prl,aps,multicol,psfig]{revtex}
\begin{document}
\input epsf

\title{ Mutual information and self-control of a fully-connected
low-activity neural network }
\author{ D.~Boll\'e }
\address{Instituut voor Theoretische Fysica, Katholieke Universiteit    
Leuven\\B-3001 Leuven, Belgium }
\author{ D.~Dominguez Carreta }
\address{ ESCET, Universidad Rey Juan Carlos, 
C/Tulipan, Mostoles, 28933 Madrid, Spain}
\date{\today}
\maketitle
\thispagestyle{empty}

\begin{abstract}
A self-control mechanism for the dynamics of a three-state fully-connected
neural network is studied through the introduction of a time-dependent  
threshold. The self-adapting threshold is a function of both the neural 
and the pattern activity in the network. The time evolution of the order
parameters is obtained on the basis of a recently developed dynamical
recursive scheme. 
In the limit of low activity the mutual information is shown to be the
relevant parameter in order to determine the retrieval quality. Due to
self-control an improvement of this mutual information content as well
as an increase of the storage capacity and an enlargement of the basins
of attraction are found. These results are compared with numerical
simulations.

\end{abstract}


\begin{multicols}{2}
\narrowtext

\section{Introduction}

It is well-known by now that low-activity neural network models have a larger 
storage capacity than the corresponding models with a mean 50\% activity
(see, e.g., \cite{MRS}).
However, this improvement is not always apparent in the basins of
attraction. Furthermore, for low activities the information content in a
single pattern is reduced. For these reasons it is argued that a neural 
activity control system is needed in the dynamics of the network in
order to keep its activity the same as the one for the memorized
patterns during the whole retrieval process \cite{Ok96}.
Recently, new suggestions have been put forward for the choice of 
threshold functions in network models in order to get an enhanced
retrieval quality-- overlap, basin of attraction, critical capacity, 
information content (see \cite{G,KA,DB98,DBA,BM} and references therein).
Diluted models \cite{G,DB98,DBA}, layered models \cite{BM} and models for
sequential patterns \cite{KA} have been considered. In all cases it has 
been found that appropriate thresholds lead to considerable improvements
of the retrieval quality.

The models mentioned above have a common property. For the diluted and
layered models there is no feedback in the dynamics. For the model with
sequential patterns no feedback correlations are taken into account. The
absence of feedback considerably simplifies the dynamics. Hence, it
is interesting to look at a model with feedback correlations
and to see whether the introduction of a threshold in the sense
described above still enhances the retrieval properties in this much
more complex situation. 
  
With these ideas in mind we consider in the sequel low activity (or in  
other words sparsely coded) fully connected neural networks. 
In particular, we study the application of a self-control mechanism 
proposed recently for a  diluted network of binary patterns \cite{DB98}.
Self-control has been introduced in order to avoid imposing some 
external constraints on the network with the purpose of improving its
retrieval properties. Such external constraints destroy the autonomous
functioning of the network.

The model we look at is a fully-connected attractor neural network
 with neurons and patterns taking the values $\{-1,0,+1\}$ 
and pattern activity $a$.
A low-activity neural network corresponds then to the case where the
pattern distribution is far from uniform, i.e., $a<2/3$. This network has the
advantage that it can be generated keeping a symmetric distribution of  
the states since both the $\pm 1$ states are considered the active ones,
while the $0$ state is the inactive one.

The rest of this paper is organised as follows. The three-state network
model and its order parameters are described in section II. In order to
study the retrieval quality of the model, especially in the limit of low
activity the mutual information content is analysed in Section III.
Section IV discusses the dynamics of this network in the presence of the 
self-control mechanism realised through the introduction of a time-dependent
threshold. Evolution equations for the order parameters are written
down. Using these equations the influence of self-control on the
retrieval quality of the
network -- information content, critical capacity, basins of attraction
-- is studied in section V.  Furthermore, these theoretical findings
are compared with results from  numerical simulations of a fully
connected network of $10^4$ neurons.
Finally, section VI presents some concluding remarks.

\section{The model}

Consider a neural network model of $N$ three-state neurons. At a
discrete time step $t$ the neurons $\sigma_i \in \{0, \pm 1\},
i=1,\ldots,N$ are updated according to the parallel deterministic
dynamics 
\begin{equation}
   \sigma_{i,t+1}= F_{\theta_{t}}(h_{i,t}),
             \quad
       h_{i,t}= \sum_{i(\neq j)}^{N}J_{ij}\sigma_{j,t}
        \label{2.si}
\end{equation}
where $h_{i,t}$ is the local field of neuron $i$ at time $t$ and
$\theta_{t}$ a time-dependent threshold parameter. 
As usual, the transfer function $F_{\theta_{t}}$ is given by
\begin{equation}
       F_{\theta_t}(x)\equiv sgn(x)\Theta(|x|-\theta_t)
       \label{2.Ft}
\end{equation}
with $\Theta$ the standard Heaviside function. 

The couplings $J_{ij}$ are determined as a function of the memorized
patterns  $\xi^{\mu}_{i}$ by the Hebbian learning algorithm 
\begin{equation}
    J_{ij}= 
       {1\over Na}\sum_{\mu=1}^{p=\alpha N} \xi^{\mu}_{i}\xi^{\mu}_{j}
          \label{2.Ji}
\end{equation}
with $\alpha$ the loading capacity. The patterns are taken to be
independent identically distributed random variables (IIDRV)
$\xi^{\mu}_{i} \in\{0,\pm 1\}, i=1, \ldots,N, \mu=1,\ldots,p$,
chosen according to the probability distribution
\begin{equation}
  p(\xi^{\mu}_{i})=
       a\delta(|\xi^{\mu}_{i}|^{2}-1)+(1-a)\delta(\xi^{\mu}_{i})
         \label{1.px}
\end{equation}
with $a=\langle|\xi^{\mu}_{i}|^{2}\rangle $ the activity of the
patterns. 
Moreover, we assume that there is no bias, i.e., 
$\langle\xi^{\mu}_{i} \rangle=0$
and that there exist no correlation between patterns such that 
$\langle\xi^{\mu}_{i}\xi^{\nu}_{i}\rangle=0$.

At this point we remark that the long-time behavior of this network
model is governed by the spin-1 Hamiltonian
\begin{equation}
     H=-\sum_{i,j}J_{ij}\sigma_i\sigma_j-\theta_t\sum_i\sigma_i^2.
        \label{2.Hi}
\end{equation}
Furthermore, the Hopfield model can be recovered by taking the
activity $a=1$ and the threshold $\theta_{t}=0$. 

The standard order parameters of this type of models are the retrieval
overlap between the $\mu$th-pattern and the microscopic state of the network
\begin{equation}
   m^{\mu}_{N,t}\equiv{1\over aN}\sum_{i}\xi^{\mu}_{i}\sigma_{i,t},
      \label{2.mm}
\end{equation}
and the neural activity of the neurons
\begin{equation}
   q_{N,t}\equiv{1\over N}\sum_{i}|\sigma_{i,t}|^{2}.
       \label{2.qN}
\end{equation}

In the next Section we use these order parameters in order to study the
retrieval quality of the network.

\section{Mutual information}

It is known that the Hamming distance between the state
of the network and the pattern $\{\xi_i^{\mu}\}$, viz.
\begin{eqnarray}
  d^{\mu}_{t}\equiv{1\over N}\sum_{i}|\xi^{\mu}_{i}-\sigma_{i,t}|^{2}
             = a-2am^{\mu}_{N,t}+q_{N,t}
        \label{2.Em}
\end{eqnarray}
is a good measure for the
retrieval quality of a network when the patterns are uniformly
distributed, i.e., when the neural activity $a=2/3$. But for
low-activity networks it cannot distinguish between a situation where
most of the wrong neurons $(\sigma_{i} \neq \xi^{\mu}_{i})$ are turned
off and a situation where these wrong neurons are turned on. This
distinction is critical in the low-activity three-state network 
because the inactive neurons carry less information than the
active ones \cite{DB98}. Therefore the mutual information 
function $I(\sigma_{i,t};\xi_{i,t}^\mu)$ has been introduced 
\cite{DB98,Bl90} 
\begin{equation}  
    \label{eq:inf}
   I(\sigma_{i,t};\xi_{i,t}^\mu)=
       S(\sigma_{i,t})-\langle S(\sigma_{i,t}|\xi_{i,t}^\mu)\rangle
                                 _{\xi^{\mu}_t}
\end{equation}
where $\xi_{i,t}^{\mu}$ is considered as the input and $\sigma_{i,t}$ as
the output with $S(\sigma_{i,t})$ its entropy and
$S(\sigma_{i,t}|\xi_{i,t}^\mu)$ its conditional entropy, viz.
\begin{eqnarray} 
      \label{eq:en}
  S(\sigma_{i,t})&=&-\sum_\sigma p(\sigma_{i,t})\ln[p(\sigma_{i,t})]\\ 
      \label{eq:enc}
  S(\sigma_{i,t}|\xi_{i,t}^\mu )&=&
       -\sum_\sigma p(\sigma_{i,t}|\xi_{i,t}^\mu)
                \ln[p(\sigma_{i,t}|\xi_{i,t}^\mu)]~.
\end{eqnarray}
Here $p(\sigma_{i,t})$ denotes the probability distribution for the
neurons at time $t$ and $p(\sigma_{i,t}|\xi_{i,t}^\mu )$ indicates the
conditional probability that the $i$-th neuron is in a state
$\sigma_{i,t}$ at time $t$ given that the $i$-th site of the stored
pattern to be  retrieved is $\xi_{i,t}^\mu$.

The calculation of the different terms of this mutual information for
the model at hand proceeds as follows.
As a consequence of the mean-field theory character of our model it is
enough to consider the distribution of a single typical neuron so we
forget about the index $i$ in the sequel. We also do not write the time
index $t$ and the pattern index $\mu$.

The conditional probability that the $ith$ neuron is in a state
$\sigma_{i}$ at time $t$, given that the $ith$ site of the pattern being
retrieved is $\xi_{i}$ can be obtained as follows. Formally writing   
$\langle O \rangle = \langle\langle O \rangle_{\sigma|\xi}\rangle_{\xi}=
\sum_{\xi} p(\xi) \sum_{\sigma} p(\sigma|\xi) O $ for an arbitrary
quantity $O$ and using the complete knowledge about the system
$\langle \xi \rangle=0, \, \langle \sigma \rangle=0, \,
\langle \sigma \xi \rangle=am, \,\langle \xi^2 \rangle=a, \,
\langle \sigma^2 \rangle=q, \, \langle \sigma^2 \xi \rangle=0, \,
\langle \sigma \xi^2 \rangle=0, \, \langle \sigma^2 \xi^2 \rangle=an,
\, \langle 1 \rangle=1$ we arrive at
\begin{eqnarray}
  p(\sigma|\xi)&&= (s_{\xi}+m\xi\sigma)\delta(\sigma^{2}-1)
                 +(1-s_{\xi})\delta(\sigma),
		 \nonumber\\
          s_{\xi}&&\equiv s-{q-n\over 1-a}\xi^{2},\,\,
          s\equiv {q-an\over 1-a}.
\label{3.ps}
\end{eqnarray}

At this point we see from (\ref{3.ps}) that besides $m$ and $q$
the following parameter 
\begin{equation}
   n^{\mu}_{N,t}\equiv 
         {1\over aN}\sum_{i}^{N}|\sigma_{i,t}|^{2}|\xi^{\mu}_{i}|^{2}
   \label{2.nm}
\end{equation}
will play an independent role in the mutual information function.
This quantity is called the activity-overlap since it determines
the overlap between the active neurons, $|\sigma_{it}|=1$, and 
the active parts of the memorized patterns, $|\xi^{\mu}_{i}|=1$.
We remark that it also shows up in the alternative expression of the
retrieval quality through the performance  $P_t^\mu=1/N \sum_i
\delta_{\xi^\mu_{i,t},\sigma_{i,t}}$ (see \cite{R90,SWB97}).
It does not play any independent role in the time evolution of
the network, independent of the architecture considered -- diluted, layered
or fully-connected.

Next, one can verify that the probability $p(\sigma|\xi)$ is consistent
with the averages
\begin{eqnarray}
   m&&=
  \frac1a\langle\langle\sigma\rangle_{\sigma|\xi}{\xi}\rangle_{\xi},
    \label{3.ma1} \\
  q&&=
  \langle\langle\sigma^{2}\rangle_{\sigma|\xi}\rangle_{\xi},
   \label{3.ma2} \\
  n&&=
   \frac1a\langle\langle\sigma^{2}\rangle_{\sigma|\xi}{\xi^{2}}
                \rangle_{\xi}. 
\label{3.ma}
\end{eqnarray}
These averages are precisely equal in the limit $N \rightarrow \infty$
to the order parameters $m$ and $q$ in eq.~(\ref{2.mm})-(\ref{2.qN}) and
to the activity-overlap defined in eq.~(\ref{2.nm}).(The fluctuations
around their mean values can be neglected according to the LLN, hence
the  average over a particular $i$-site distribution equals the infinite
sum over $i$).

Using the probability distribution of the memorized patterns
(\ref{1.px}) we furthermore obtain
\begin{equation}
    p(\sigma)\equiv\sum_{\xi}p(\xi)p(\sigma|\xi)=
             q\delta(\sigma^{2}-1)+(1-q)\delta(\sigma).
   \label{3.px}
\end{equation}

The expressions for the entropies defined above then become 
\begin{eqnarray}
   &&S(\sigma)= - q\ln{q\over 2} - (1-q)\ln(1-q),
	               \\
   &&\langle S(\sigma|\xi)\rangle_{\xi}=
                  a S_{a} + (1-a) S_{1-a},  
		        \\
  &&S_{a}=-{n+m\over 2}\ln{n+m\over 2}-{n-m\over 2}\ln{n-m\over 2}  
                    \nonumber \\
                             &&-(1-n)\ln(1-n), \\
    &&S_{1-a}= - s\ln{s\over 2} - (1-s)\ln(1-s)
      \label{3.Hs}
\end{eqnarray}
and the mutual information is then given by eq.~(\ref{eq:inf}). We
recall that $m_t$, $q_t$ as well as $n_t$ are needed in order to
completely know the mutual information content of the network at time
$t$.

\section{Threshold Dynamics}

It is known that the parallel dynamics of fully connected networks is
difficult to solve, even at zero temperature, because of the strong
feedback correlations \cite{BKS}. Recently, a recursive dynamical scheme
has been
developed which calculates the distribution of the local field at a
general time step using signal-to-noise analysis techniques
\cite{BJS98}. Recursion relations are obtained determining the
full time evolution of the order parameters. We shortly review these
results.

Suppose that the initial configuration of the network
$\{\sigma_{i,0}\}$, is a collection of IIDRV with mean
$\left\langle\sigma_{i,0}\right\rangle=0$, variance
$\left\langle(\sigma_{i,0})^2\right\rangle=q_{in}$, and
correlated  with
only one stored pattern, say the first one $\{\xi^1_i\}$:
\begin{equation}
        \label{eq:init1}
        \frac{1}{N}\sum_i\xi_i^\mu\sigma_{i,0}=
          \delta_{\mu,1}m^1_{in}a\,,    \quad m^1_{in}>0 \, .
\end{equation}
This implies that by the law of large numbers (LLN) one gets for the
retrieval overlap and the activity at $t=0$
\begin{eqnarray}
        m^1_0&\equiv&\lim_{N \rightarrow \infty} m^1_{N,0}
                =\frac1a \left\langle\xi^1_i \sigma_{i,0}\right\rangle
                = m^1_{in}                                          \\
        q_0&\equiv&\lim_{N \rightarrow \infty} q_{N,0}
                = \left\langle\sigma_{i,0}^2\right\rangle=q_{in}\,.
\end{eqnarray}
At a given time step of the dynamics, the state of a neuron,
$\sigma_{i,t+1}$ is determined by its local field at the previous time
step.
In general, in the limit $N \rightarrow \infty$  the distribution of the
local field at time $t+1$ consists
out of a discrete part and a normally distributed part \cite{BJS98}
\begin{eqnarray}
       && h_{i,t}=\xi _i^1 m^1_t +\sqrt{\alpha a D_t}\, {\cal N}(0,1)    
	     + B_{i,t} 
	\label{eq:init2} \\
       && D_t=\mbox{Var}\left[ r^\nu_t\right]
	     = \mbox{Var}\left[\lim_{ N \to \infty} \frac{1}{a \sqrt{N}}  
	    \sum_{i} \xi_i^{\nu} \sigma_{i,t}\right],  \nu >1
	   \label{eq:DVar}  \\
       && B_{i,t}= \sum_{t'=0}^{t-1} \alpha
         \left[\prod_{s=t'}^{t-1} \chi_s\right] \, \sigma _{i,t'}  
         \label{eq:MM} 
\end{eqnarray}			
with ${\cal N}(0,1)$ a Gaussian random variable with mean zero and
variance $1$ and $\chi$ the susceptibility
\begin{equation}
   \chi_t= \frac{1}{\sqrt{\alpha a D_t}}    
                \left\langle\!\left\langle\int {\cal D}  z ~  z \,
                    F_{\theta_t}
                        \left( \xi^1m^1_t + \sqrt{\alpha a D_t}\,z
                        \right)\right\rangle\!\right\rangle 
	\label{eq:chifix}		
\end{equation}	
where ${\cal D}$ is the Gaussian measure. 
In the above $\langle\!\langle \cdot \rangle\!\rangle$
denotes the average both over the distribution of the embedded patterns
$\{\xi_i^\mu\}$ and the initial configurations $\{\sigma_{i,0}\}$. The
average over the initial configurations is hidden in an average over the
local field through the updating rule (\ref{2.si}).

The first term on the  r.h.s. of (\ref{eq:init2}) is the signal term 
produced by the pattern that is being retrieved, the rest represents the
noise induced by the $(p-1)$ non-condensed patterns. In particular, the 
second term is Gaussian noise and the last term $B_{i,t}$ contains  
discrete noise coming from the feedback correlations. The quantity $D_t$
satisfies the recursion relation
\begin{equation}
        \label{eq:Drec}
        D_{t+1}=\frac{q_{t+1}}{a}+\chi^2_t D_t +
                2 \chi_t \mbox{Cov}[{\cal N}(0, q_{t+1}/a),r^{\mu}_t] 
\end{equation}
For more details we refer to \cite{BJS98}. Using the above scheme
the order parameters at a general time step can then be obtained in the
limit $N \to \infty$ from Eqs.~(\ref{2.mm})-(\ref{2.qN}) and (\ref{2.si}) 
\begin{eqnarray}
        \label{eq:m}
        m^1_{t+1} &=& \frac{1}{a} \langle\!\langle
                 \xi_i^1 F_{\theta_t}(h_{i,t}) \rangle\!\rangle      \\
        \label{eq:a}
        q_{t+1}   &=& \langle\!\langle F_{\theta_t}^2(h_{i,t})
                         \rangle\!\rangle \, .
\end{eqnarray}
The activity overlap needed in order to find the mutual information can
also be written as
\begin{equation}
  \label{eq:newn}
        n^1_{t+1} = \frac{1}{a} \langle\!\langle
                 (\xi_i^1)^2 F_{\theta_t}^2(h_{i,t}) \rangle\!\rangle
		      \,. 
\end{equation}
Of course, we then also need to specify its initial value
\begin{equation}
    n^1_0\equiv\lim_{N \rightarrow \infty} n^1_{N,0}
         =\frac1a \left\langle(\xi^1_i)^2 (\sigma_{i,0})^2\right\rangle
	         \,. 
\end{equation}

The idea of the self-control threshold dynamics introduced in the
diluted model \cite{DB98} and studied for some other models without
feedback correlations \cite{KA,DBA,BM} has been  precisely to let the
network counter the noise term in the local field at each step of the
dynamics by introducing the following form  
\begin{equation}
   \theta_{t}=c(a)\sqrt{\alpha a D_{t}},
    \label{2.tt}
\end{equation}
where the function $c(a)$ is a function of the pattern activity.
We remark that in these cases without feedback there is no discrete
noise in the local field (the term $B_{i,t}$ in eq.~(\ref{eq:init2}) is 
absent). Furthermore, also the covariance term in eq.~(\ref{eq:Drec}) is
absent. Moreover, for the diluted model $D_t= q_{t}/a$, i.e., only the
first term in eq.~(\ref{eq:Drec}) is present. So this dynamical
threshold has two important characteristics. First, it is a macroscopic 
parameter having the same value for every neuron thus no average must be
taken  over the microscopic random variables at each time step.
Secondly, it changes each time step but no statistical history
intervenes in this proces.

We see that the choice (\ref{2.tt}) is in fact related to the variance
of the local field, taken for a fixed realization of the pattern which
is being retrieved. It is the width of the noise produced by the
non-condensed patterns. 
It is obvious that it cannot be taken to be a function of the overlap
with the pattern being retrieved. 
As a consequence, for the fully connected network we cannot work with
the exact form for $D_t$ as given in eq.~(\ref{eq:Drec}) because of the
presence of the covariance. So, if we want to take into account some
effects of feedback correlations and if we want the threshold to have
the characteristic properties mentioned above, we need to approximate
the covariance 
term in eq.~(\ref{eq:Drec}) such that only the previous time step is
involved. This is realised by approximating this term by  
$2 \chi_t \{\mbox{Var}[{\cal N}(0,q_{t+1}/a)]\mbox{Var}[r_t^\mu]\}^{1/2}
 = 2\chi_t [q_{t+1}/a]^{1/2} [D_t]^{1/2}$. We then easily get 
\begin{eqnarray} 
   \alpha a D_{t+1}&=& \left[ G_{t}+\sqrt{\alpha q_{t+1}} 
                    \right]^2 
   \label{4.Va0} \\
         G_t&=&\left\langle\!\left\langle\int {\cal D}  z ~  z \,
                    F_{\theta_t}
                        \left( \xi^1m^1_t + \sqrt{\alpha a D_t}\,z
                        \right)\right\rangle\!\right\rangle 
    \label{4.Va}
\end{eqnarray}
Furthermore, we take both contributions at equal times and call 
$\sqrt{\alpha a D_t}\equiv \Delta_t=G_t+\sqrt{\alpha q_t}$. 
For more details on this approximation of the feedback
correlations we refer to \cite{BJS98,DT95} and references therein.
Finally, since $G_t$ is a function of the overlap $m_t^1$, a quantity
which is not available to the network we replace it by
$G_0=\sqrt{2/\pi}\,a$. 

What is left then is to find a form for $c(a)$.
For the low-activity networks considered up to now 
the storage capacity could be considerably improved by taking $c(a)=    
[-2\ln(a)]^{1/2}$ such that for the diluted model
$\theta_{t}=[-2\ln(a)\alpha q_t]^{-1/2}$.
The same form has been shown to work for the layered model \cite{BM}.
For the fully connected model considered here we again propose, a
priori, this form.
So, combining these results we take as self-control threshold 
\begin{eqnarray}
         \theta_{t}&=&\sqrt{-2\ln(a)}\,\,\Delta^{0}_{t}\,
       \label{4.tt}   \\
        \Delta^{0}_{t}&=& \sqrt{2/\pi}\,a + \sqrt{\alpha q_{t}}.
          \label{4.Dt}
\end{eqnarray}

Finally, we make one more assumption on the dynamics. In the
local field distribution (\ref{eq:init2}) we forget about the discrete
noise $B_{i,t}$ and suppose  that the
noise produced by the non-condensed patterns is Gaussian distributed.
Computer simulations have shown that this assumption is approximately
valid as long as the retrieval is succesful \cite{NO93}. As a
consequence we can write down recursion relations for the order
parameters 
\begin{eqnarray}
  m_{t+1}&=& \int {\cal D} z F_{\theta_{t}}(m^1_{t}+z\Delta_{t})
         \label{4.mt}         \\
  q_{t+1} &=& a \int {\cal D} z
                [F_{\theta_{t}}(m^1_{t}+z\Delta_{t})]^{2}
		\nonumber \\
          && +(1-a)  \int {\cal D} z [F_{\theta_{t}}(z\Delta_{t})]^{2}
   \label{4.qt}
\end{eqnarray}   
with 
\begin{eqnarray}  
   \Delta_{t}&=& \sqrt{\alpha q_{t}} +
       a \int {\cal D}  z ~  z \,
                    F_{\theta_{t}}( m^1_{t} +  z\Delta_{t} ) \nonumber\\
        && +(1-a) \int {\cal D}  z ~  z \,
                    F_{\theta_{t}} ( z\Delta_{t})
      \label{4.deltat}
\end{eqnarray}
where we have already averaged over $\xi$. We remark that the form of
the last equation is different from the corresponding equation for the
diluted and
the layered versions of this model \cite{BM,BJSrev} because of the feedback.

The expressions for the overlap $m_{t+1}$, the neural activity $q_{t+1}$
and the noise $\Delta_{t}$ due to the non-condensed patterns 
describe the (approximate) macro-dynamics of the fully-connected 
neural network. Besides the self-control model with the threshold given
by eq.~(\ref{4.tt})-(\ref{4.Dt}) we also consider the
model with the threshold fixed at its zero time value, i.e.,
$\theta_{t}=\theta_{0}$. 

At this point we remark that when studying the mutual information, we
want to introduce explicitly the
activity-overlap parameter, $ n_t$ (recall eq.~(\ref{2.nm})) in the dynamics
leading to the following expression for $q_{t+1}$
\begin{equation}
  q_{t+1}= a\,n_{t+1} + (1-a)\, s_{t+1}
  \label{qas}
\end{equation}
where $s_{t+1}$ is then, obviously, defined by the integral of 
$[F_{\theta_{t}}(z\Delta_{t})]^{2}$. This parameter $s$ is precisely
that introduced in Eq.~(\ref{3.ps}) and measures the number of active
neurons in inactive condensed pattern sites.

In the following section we compare the retrieval properties of a fully
connected network governed by this approximate dynamics with and without
self-control with numerical simulations. The main aim is to show that   
self-control also works in the case of fully connected models.

\section{Results}

\subsection{Numerical Results}

We have solved the time evolution of our threshold dynamics with a
time-dependent self-control threshold given by
eq.~(\ref{4.tt})-(\ref{4.Dt})) and with a time-independent threshold
where the neural activity is fixed  by $q_t=q_0$. 
\vspace{-2.5cm}
\begin{figure}[t]
\begin{center}
\epsfysize=11cm
\leavevmode
\epsfbox[1 1 700 700]{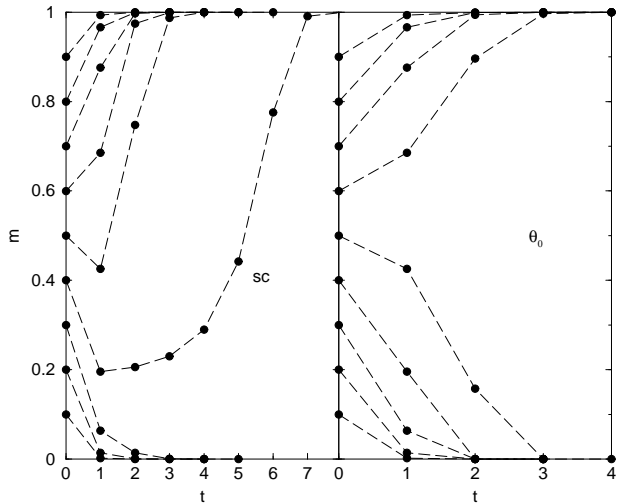}
\vspace{-2cm}
\end{center}
\caption{ {\em The evolution of the overlap $m_t$ for several initial
values $m_0$, with $q_0=0.01=a$ and $\alpha=2$ for the self-control
model (left) and the fixed-threshold model (right).
The dashed curves are a guide to the eye.} }
\label{I,tA}
\end{figure}
We have studied the behavior of these networks in the
range of pattern activities $0.01\leq a\leq 0.67$ i.e., from
low-activities to a uniform distribution of patterns.

For both thresholds it turned out that the best results were
obtained by taking $c(a)=\sqrt{-2\ln(a)}+K$, with $K=0$ for $a\geq 0.1$
and $K=0.5$ for $a<0.1$. At this point we remark, however, that the pure
log form for $c(a)$ is derived in the theoretical limit $a \rightarrow
0$. So, it may be that we did not reach small enough values in our
numerical analysis (which is due to numerical complexity). We recall
that one of the main aims of this work is to show that self-control also
works for fully connected models. 
\vspace{-1cm}
\begin{figure}[b]
\begin{center}
\epsfysize=10cm
\leavevmode
\epsfbox[1 1 700 700]{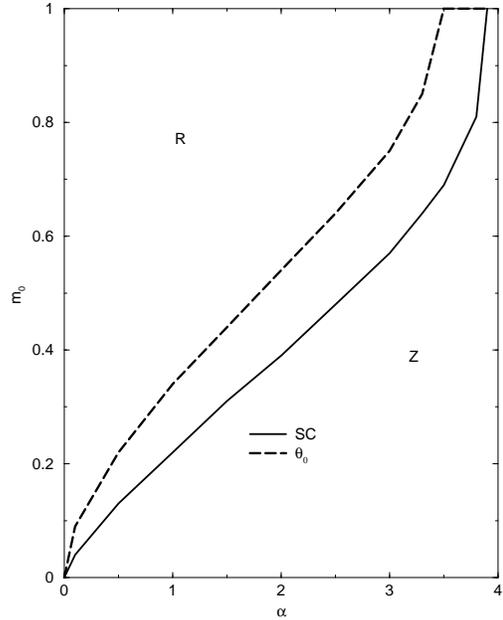}
\vspace{-1.5cm}
\end{center}
\caption{ {\em The basin of attraction for the retrieval overlap $m_0$ as a
function of $\alpha$ for
$a=0.01$ and initial $q_0=a$ and $n_0=1$ for the self-control model
(full line) and the fixed threshold model (dashed line). } }
\label{m,tA}
\end{figure}

The important features of self-control are illustrated in Figs.~1-5. In
Fig.~1 we compare the time evolution of the retrieval overlap, $m_t$,
starting from several initial values, $m_0$, for the model with
self-control, $\theta_{sc}=\theta_{t}$ (recall
eq.~(\ref{4.tt})-(\ref{4.Dt}))  with the model with fixed threshold
$\theta_{0}$. An initial neural activity $q_0=a=0.01$ and a loading
$\alpha=2$ have been taken. 
We observe that the self-control forces more of the
overlap trajectories to go to the retrieval attractor $m=1$. Only an
initial overlap $m_0 \sim 0.4$ for the self-control model versus $m_0
\sim 0.6$ for the fixed threshold model is needed.
We remark that for $m_0\leq 0.6$  the overlap decreases in the first
time step for both models. This is an expression of the fact that
correlations are especially important in the first time steps leading to
a decreasing neural activity $q_t$, but the self-control threshold is
able to counter these effects. Near the attractor correlations seem to
become less important and the Gaussian character of the local field
distribution dominates.    

Since the initial overlap needed to retrieve a pattern is smaller for
the self-control model, the basins of attraction of the patterns are
substantially larger. This is further illustrated in Figs.~2-4 where the
basin of atraction for the whole retrieval phase $R$ is shown for both
models with an initial value $q_0=a=0.01$. We have calculated the
fixed-point $m_{\infty}$ of the dynamics (\ref{4.mt}), (\ref{4.qt}) and 
(\ref{4.deltat}) and we have determined the intial conditions of the
relevant parameters such that the network is able to retrieve, i.e.,
such that $m_\infty\sim 1$. It is interesting to also give $n_0$ and/or
$s_0$ separately in order to see how the activity $q_t$ is built up.
\vspace{-1.8cm}
\begin{figure}[b]
\begin{center}
\epsfysize=11cm
\leavevmode
\epsfbox[1 1 700 700]{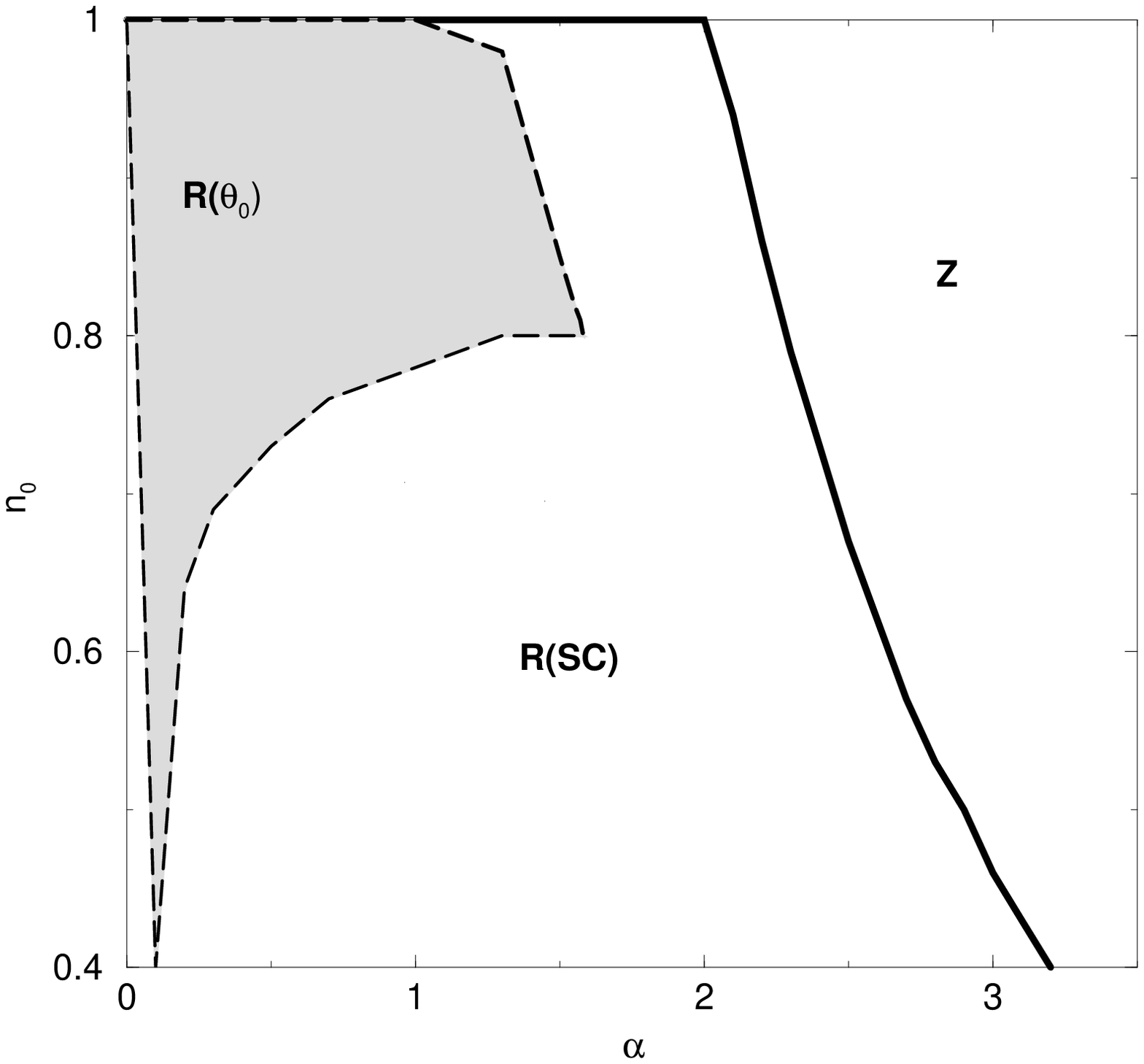}
\vspace{-3cm}
\end{center}
\caption{ {\em  The basin of attraction for the activity-ovelap $n_0$
as a function of $\alpha$ for $a=0.01$ and initial $q_0=an_0$ and
$m_0=0.4$ for the self-control model (full line) and the fixed threshold
model (dashed line). } }
\label{m0,a}
\end{figure}
In Fig.~2 we have used $q_0=a, n_0=1$. The basin of attraction for the
self-control model is larger, even near the border of critical storage.
Hence the storage capacity itself is also bigger. 
\vspace{-1cm}
\begin{figure}[h]
\begin{center}
\epsfysize=10cm
\leavevmode
\epsfbox[1 1 700 700]{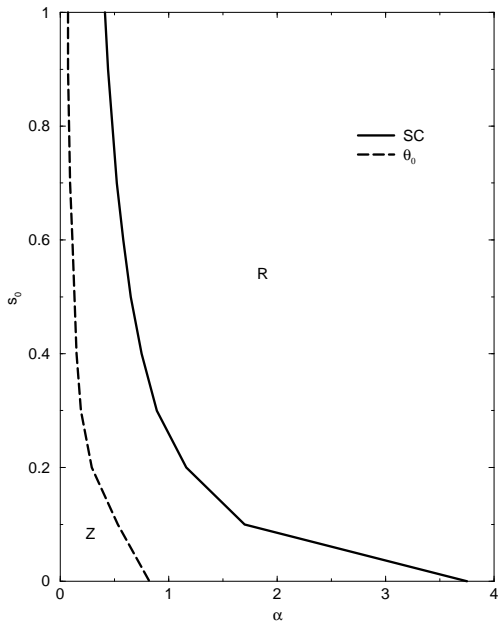}
\vspace{-1.5cm}
\end{center}
\caption{ {\em The basin of attraction for the parameter $s_0$ (see main
text) as a function of $\alpha$ for $a=0.01$  with initial $m_0=1$ and
$n_0=1$ for the self-control model (full line) and the fixed threshold
model (dashed line). } }
\label{Ii,a}
\end{figure}
Furthermore, a smaller
initial activity-overlap $n_0$ suffices to have retrieval as is 
seen in Fig.~3.
There we start with initial conditions $m_0=0.4$, i.e., the smallest
initial overlap possible for $\alpha=2$ as we recall from Fig.~1, and
$q_0=an_0$  or,
equivalently $s_0=0$. So we consider small $q_0$ running from $0.004$ to
$0.01$.  
We observe the peculiar behavior that for the fixed-threshold network an
initial $n_0>m_0$ is needed, but even then still no retrieval is
possible for low storage $\alpha<0.1$. For the self-control model a much
broader region of retrieval exists.
Finally, the specific role of the parameter $s_t$ is displayed in Fig.~4.
We start from a maximal initial overlap $m_0=1$ and take $n_0=1$ meaning
that for $s_0=0$ to $s_0=1$, $q_t$ runs from $0.01$ to $1$. It can be
seen that especially when $s_0$ is getting large the storage capacity of
both models decreases quite drastically but again much less for the
self-control model.  
%\vspace{-0.5cm}
\begin{figure}[h]
\begin{center}
\epsfysize=10cm
\leavevmode
\epsfbox[1 1 700 700]{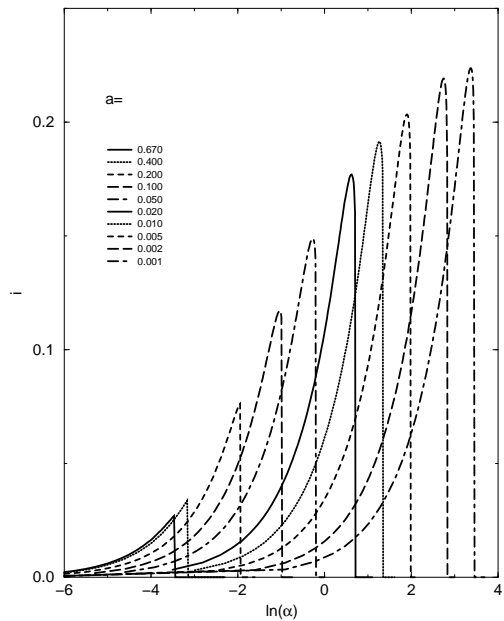}
%\vspace{-1cm}
\end{center}
\caption{ {\em The information $i$ as a function of
$\alpha$ with self-control for several values of $a$: results using the
evolution equations.} }
\label{ia,A}
\end{figure}

We conclude with the observation that self-control works in a large
range of pattern activities $a$, as shown in Fig.~5. There the mutual   
information content $i= \sum_i \sum_{\mu} I/(\#J_{ij} )=\alpha
I$ is  plotted as a function of
the loading $\alpha$ on a logaritmic scale.
We observe the slow increase of $i$ as the activity $a$
decreases, saturating at a value close to $ i\sim 0.3$. This
behavior is typical for low activity networks \cite{Pe89,Ho89}.

\subsection{Simulations}
Simulations have been carried out for systems with $N=10^4$ neurons.
For every new stored pattern $\mu$, we start our dynamics with a state  
$\sigma_{i}=\xi^{\mu}_i$, and calculate the order parameters $m_t$,
$n_t$ and the activity overlap $q_t$, using the definitions (\ref{2.mm}),
(\ref{2.qN}) and (\ref{2.nm}).
To avoid a very large computation time we have stopped the dynamics
after $t\leq 5$ time steps when no convergence was reached before.
Then we have averaged over windows in the $p$-axis in order to obtain
the  mutual information $i$.
The window size runs from $\delta p=50$ for $a=0.67$ (where we have
stored $p=10^3$ patterns) up to $\delta p=2\times 10^3$ for $a=0.01$
(where we have stored $p=5\times 10^4$ patterns).
\vspace{-0.5cm}
\begin{figure}[h]
\begin{center}
\epsfysize=10cm
\leavevmode
\epsfbox[1 1 700 700]{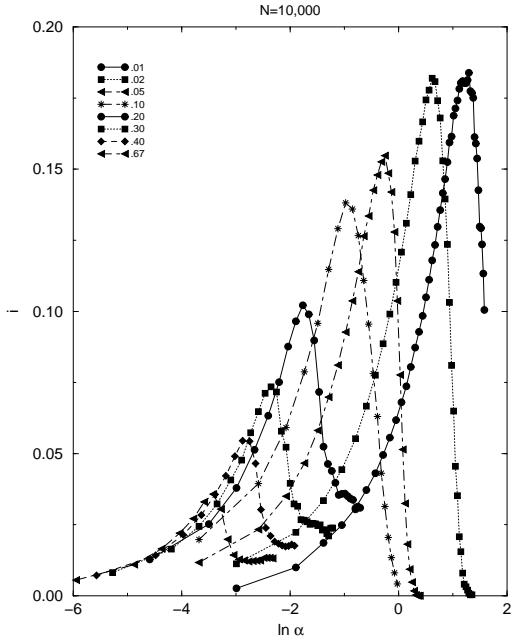}
\vspace{-1cm}
\end{center}
\caption{ {\em The information $i$ as a function of
$\alpha$ with self-control for several values of $a$:
simulations with $N=10^4$ neurons. } }
\label{im,M}
\end{figure}

The conditions on the LLN mentioned in Section IVA are approximatelly
fullfilled for such large networks, since the fluctuations 
(neglected in Eqs.(\ref{3.ma1})-(\ref{3.ma})) are of order 
$1/\sqrt{aN}$.
However, for smaller activities $a$, this quantity may be not so small.
It becomes crucial in the case $a=0.01$, where this quantity is $0.1$,
such that the finite size effects get relevant. This implies a kind of
cut-off in the information for the self-control model as seen in Fig.~6.
However, the agreement
with the analytic results of Fig.~5 is quite good up to $a=0.01$.

In order to further understand the details of the retrieval quality 
we plot in Fig.~7 all the parameters $m,n,q$ for the model with and
without self-control in two cases: $a=0.01$, a low-activity case, 
and $a=0.67$, implying a uniform distribution of patterns.
For the uniform case we do not see a big  difference between the two
models (self-control and fixed threshold). Only for larger values of
$\alpha$, self-control shows a little improvement. 
For the low-activity case, however, the main role of self-control on the
neural activity is clearly noticed since $q\sim a$ in that case while in
the fixed-threshold model it is impossible to control $q$ such that it
stays in the neigborhood of $a$. As a conseqeunce the mutual
information, e.g., is only about half of that for the model with
self-control. 
\vspace{-0.5cm}
\begin{figure}[h]
\begin{center}
\epsfysize=10cm
\leavevmode
\epsfbox[1 1 700 700]{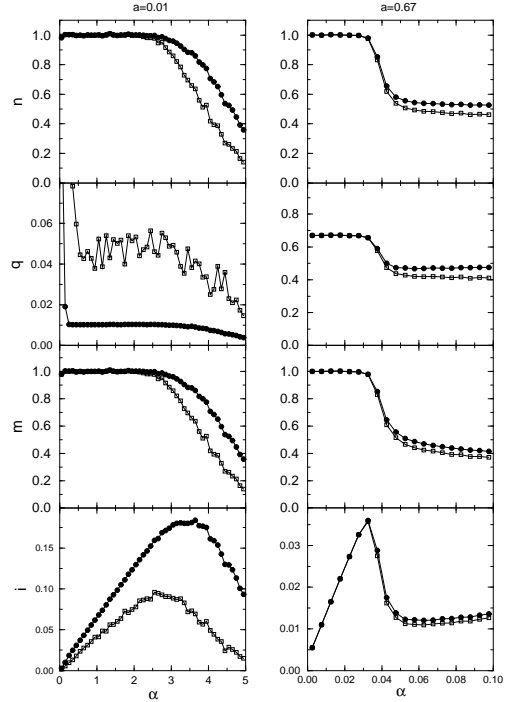}
\vspace{-4cm}
\end{center}
\caption{ {\em Simulations for $N=10^4$ neurons of the information
$i$, the overlap $m$, the neural
activity  $q$ and the activity-overlap $n$ as a function of $\alpha$ for
the self-control model (filled circles) and the fixed threshold model
(squares) after five time steps. The pattern activities are
$a=0.01$ (left) and $a=0.67$ (right). } }
\label{ia1,A}
\end{figure}

Finally, in Fig.~8 we compare the simulations with the results from
the fixed-points of the dynamics (\ref{4.mt}), (\ref{4.qt}) and
(\ref{4.deltat}) for $a=0.03$. Up to $t=5$ time steps are considered for
both the model with and without self-control and we have averaged over a
window in the $p$-axis of size $\delta p=10^3$. 
For the self-control model the small underestimation of the theoretical
results can, of course,  be attributed to the approximations
of the noise term (recall Eqs~(\ref{4.Va0}) and (\ref{4.Dt})).
\vspace{-0.5cm}
\begin{figure}[h]
\begin{center}
\epsfysize=10cm
\leavevmode
\epsfbox[1 1 600 800]{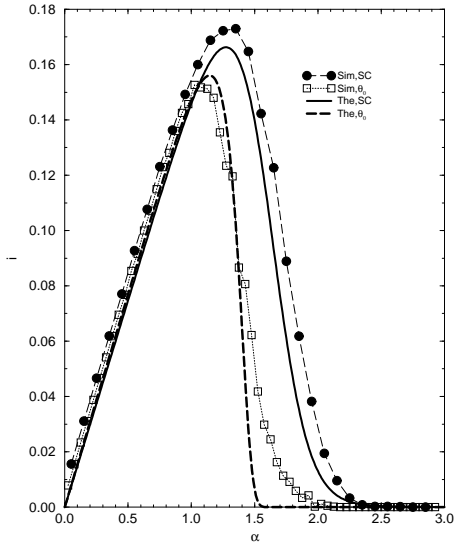}
\vspace{-1cm}
\end{center}
\caption{ {\em The information $i$ as a function of
$\alpha$ with (thick full line and filled circles) and without (thick
dashed line and squares) self-control for $a=0.03$. Comparision
between the results from the evolution equations and simulations with
$N=10^4$  neurons. }} 
\label{ia2,A}
\end{figure}

\section{Concluding Remarks}

In this paper we have introduced a self-control threshold in the
dynamics of fully connected networks with three-state neurons. This
leads to a large improvement of the quality of retrieval of the network.
The relevant quantity in order to study this, especially in the limit of
low activity is the mutual information function.
The mutual information content of the network as well as the critical
capacity and the
basins of attraction of the retrieval solutions for three-state patterns
are shown to be larger because of the self-control mechanism. 
Furthermore, since the mutual information saturates, the critical
capacity of the low-activity network behaves as
$\alpha_{c}=O(|a\ln(a)|^{-1})$. 
Numerical simulations confirm these results.

This idea of self-control might be relevant for
various dynamical systems, e.g., when trying to enlarge the basins of
attraction and convergence times. Indeed, it has been shown to work also
for both diluted and layered networks. Binary as well as ternary neurons
and patterns have been treated.
In all cases, it turns out that in the low-activity regime the
self-control threshold can be taken to be  proportional to the square
root of the neural activity of the network.

\section*{Acknowledgments}

We would like to thank G.~Jongen for useful discussions. 
This work has been supported by the Research Fund of the K.U.Leuven 
(grant OT/94/9).
One of us (D.B.) is indebted to the Fund for Scientific Research - Flanders
(Belgium) for financial support.

\end{multicols}
\end{document}